\PassOptionsToPackage{cmyk,table}{xcolor}

\makeatletter
\let\blx@rerun@biber\relax
\makeatother

\documentclass[conference,flushend]{iaria}

\usepackage[babel=true,english=american]{csquotes}
\usepackage[british]{babel}
\usepackage[shortcuts]{extdash}

\usepackage[alwaysadjust]{paralist}

\usepackage[final]{microtype}

\usepackage{relsize}
\usepackage{xurl}
\usepackage{pgfplots}
\usepackage{subcaption}
\pgfplotsset{compat=1.18}
\usepackage{amsmath}
\usepackage{balance}
\usepackage{placeins}
\usepackage{tikz}
\usepackage{booktabs}
\usepackage{fontawesome5}

\usepackage{hyperref}
\usepackage[nolist]{acronym}
\begin{acronym}[]
      \acro{api}[API]{Application Programming Interface}
      \acro{mfa}[MFA]{Multi Factor Authentication}
      \acro{sms}[SMS]{Short Message Service}
      \acro{iv}[IV]{Initialization Vector}
      \acro{rsa}[RSA]{Rivest–Shamir–Adleman}
      \acro{ctap}[CTAP]{Client To Authenticator Protocol}
      \acro{fido}[FIDO2]{Fast Identity Online}
      \acro{rp}[RP]{Relying Party}
      \acro{uaf}[UAF]{Universal Authentication Framework}
      \acro{u2f}[U2F]{Universal Second Factor}
      \acro{sha}[SHA]{Secure Hash Algorithms}
      \acro{mac}[MAC]{Message Authentication Code}
      \acro{aitm}[AITM]{Adversary in the Middle}
      \acro{mac}[MAC]{Message Authentication Code}
      \acro{hsk}[HSK]{Hardware Security Key}
      \acro{ecdh}[ECDH]{Elliptic Curve Diffie Hellman}
      \acro{ecdsa}[ECDSA]{Elliptic Curve Digital Signature Algorithm}
      \acro{rsa}[RSA]{Rivest–Shamir–Adleman}
      \acro{ed25519}[Ed25519]{Edwards-curve 25519 Digital Signature Algorithm}
      \acro{ca}[CA]{Certificate Authority}
      \acro{san}[SAN]{Subject Alternative Name}
      \acro{cn}[CN]{Common Name}
      \acro{apwg}[APWG]{Anti-Phishing Working Group}
      \acro{dns}[DNS]{Domain Name System}
\end{acronym}

\title{An Analysis of Attack Vectors Against FIDO2 Authentication}

\author{%
    \large Alexander Berladskyy and Andreas Aßmuth\,\orcidlink{0009-0002-2081-2455}\\[0.3ex]\normalsize\normalfont
    Kiel University of Applied Sciences, Kiel, Germany\\
    e-mail: {\tt alexander.berladskyy@stu.haw-kiel.de, andreas.assmuth@haw-kiel.de}
}

\newcommand*\circled[1]{\tikz[baseline=(char.base)]{
    \node[fill=white,shape=circle,draw,inner sep=1pt] (char) {#1};}}

\usepackage{ifpdf}
\ifpdf
\pdfoutput=1 
\pdftrue
\fi

\makeatletter
\def\ps@IEEEtitlepagestyle{
	\def\@oddfoot{\mycopyrightnotice}
	\def\@evenfoot{}
}
\def\mycopyrightnotice{
	{\footnotesize
		\begin{minipage}{0.8\textwidth}
			\centering
			Please cite as: Alexander Berladskyy and Andreas Aßmuth, ``An Analysis of Attack Vectors Against FIDO2 Authentication,'' in \emph{Proc of the First International Conference on Cross-Domain Security in Distributed, Intelligent and Critical Systems (CROSS-SEC 2026), Lisbon, Portugal, pp.~77--83, April 2026.}
		\end{minipage}
	}
}
\makeatother

\begin{document}

\maketitle

\begin{abstract}
    Phishing attacks remain one of the most prevalent threats to online security, with the Anti-Phishing Working Group reporting over 890,000 attacks in Q3 2025 alone. Traditional password-based authentication is particularly vulnerable to such attacks, prompting the development of more secure alternatives. This paper examines passkeys, also known as FIDO2, which claim to provide phishing-resistant authentication through asymmetric cryptography. In this approach, a private key is stored on a user's device, the authenticator, while the server stores the corresponding public key. During authentication, the server generates a challenge that the user signs with the private key; the server then verifies the signature and establishes a session. We present passkey workflows and review state-of-the-art attack vectors from related work alongside newly identified approaches. Two attacks are implemented and evaluated: the Infected Authenticator attack, which generates attacker-known keys on a corrupted authenticator, and the Authenticator Deception attack, which spoofs a target website by modifying the browser's certificate authority store, installing a valid certificate, and intercepting user traffic. An attacker relays a legitimate challenge from the real server to a user, who signs it, allowing the attacker to authenticate as the victim. Our results demonstrate that successful attacks on passkeys require substantial effort and resources. The claim that passkeys are phishing-resistant largely holds true, significantly raising the bar compared to traditional password-based authentication.
\end{abstract}


\section{Introduction}
Passwords have been the predominant authentication method for decades. A password is a user-chosen string that is hashed and stored server-side. Upon authentication, the user’s input is hashed and compared with the stored hash to verify its validity. This method is simple to implement and remains the most widely used authentication mechanism.\par 
However, passwords have significant drawbacks. They are vulnerable to phishing attacks, and in the event of a server-side data breach, the hashed passwords can be exposed and potentially cracked. The \ac{apwg} reported $892{,}494$ phishing attacks in the third quarter of 2025 alone~\cite{phishing_report}, and data breaches continue to be a prevalent threat.\par 
In response to these vulnerabilities, passkeys, based on the \ac{fido} standard, provide a more secure alternative. These authentication methods are designed to be resistant to phishing, ensuring that even if a data leak occurs, no sensitive information that could compromise the user’s security is exposed.\par 
This work investigates various attack vectors targeting \ac{fido} to assess the feasibility and cost of potential attacks, and to evaluate whether it is truly resistant to phishing. The structure of the paper is as follows: \autoref{sec:background} provides an overview of how \ac{fido} works, \autoref{sec:related_work} reviews previous research on attacks against passkeys, \autoref{sec:methodology} outlines the attack vectors examined, \autoref{sec:experiments} demonstrates two specific attacks, \autoref{sec:results_and_discussion} discusses the findings, and \autoref{sec:Conclusion_and_Future_Work} presents the conclusions and directions for future research.

\section{Background}\label{sec:background}
The following sections describe the concepts of \ac{fido} and introduce the threat model.

\subsection{Client To Authenticator Protocol}
The \ac{ctap}~\cite{ctap, fido_security} is a communication protocol that allows a client, such as a browser, to communicate with an authenticator. An authenticator can be a hardware device, like a phone or a hardware key~\cite{yubico} or software-based authenticators that communicate using the same protocol. Examples include Windows Hello or applications that implement security mechanisms. \ac{ctap}, a core component of \ac{fido}~(cf. project's website~\cite{fido2026}), enables passwordless and secure authentication on the Internet. Unlike password-based systems where authentication occurs server-side through hash comparison, authentication with \ac{ctap} is performed on the devices which belong to the user. After the authentication on the local device, e.g., by fingerprint or a PIN, the secret key can be accessed and used for authentication by signing a challenge received by the \ac{rp}. The \ac{rp} is the service which a user is trying to authenticate to. This could be any web service, providing a way to authenticate to a user account. The \ac{rp} saves the public key of the owner and thus can verify if the person is the real owner of the key. The private key, which is generated by the authenticator, remains inaccessible to the end users~-- they only need a way to unlock the authenticator, e.g., a PIN or a fingerprint, and a device that is able to use \ac{ctap}. A key pair is also called \textit{passkey} in this context.

\subsection{WebAuthn}
The second part of \ac{fido} is WebAuthn~\cite{webauthn,mozillaWebAuthn2026} (Web Authentication API). \ac{ctap} covered the communication between the client and the authenticator. The communication between the client and the \ac{rp} is done by WebAuthn. WebAuthn implements two workflows, also called ceremonies, namely signup or registration and the login or authentication workflow. The signup process begins by the user choosing a passkey login instead of a password login. The \ac{rp} receives the request, containing user information, and generates a challenge, which is random data, e.g., an array of random numbers. The challenge and the user information is stored temporarily on the server, until the ceremony is finished. Then, the server responds with the challenge, user information and server information back to the client. Now the client tries to create the necessary credentials (public-key pair), by first authenticating the user via \ac{ctap}. After the generation of the keys, the private key is used to sign the challenge~-- this process is also called assertion. Attestation on the other hand, is a certificate, verifiable by the \ac{rp}, that proves the key being generated by a specific authenticator device like a \ac{hsk}. The signed challenge, user information, server (\ac{rp}) information, device information and the public key are sent to the server. Finally, upon verifying the challenge and comparing the user information the process is done.\par
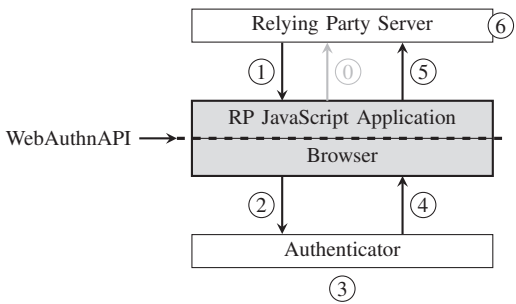
\begin{figure}[htb]
    \centering
    \begin{tikzpicture}
        \footnotesize
        \node[draw, minimum width=4cm] (rps) at (4.4, -0.5) {Relying Party Server};
        \draw[thick, fill=black!15] (2.4, -1.5) rectangle (6.4, -2.5);
        \node at (4.4, -1.75) {RP JavaScript Application};
        \draw (2.4, -2) -- (6.4, -2);
        \draw[very thick, dashed] (2.2, -2) -- (6.6, -2);
        \node at (4.4, -2.25) {Browser\vphantom{j}};
        \node[draw, minimum width=4cm] (auth) at (4.4, -3.5) {Authenticator\vphantom{j}};
        
        \draw[thick, stealth-] (2.2, -2) -- ++(-0.5, 0) node[left]{WebAuthnAPI};

        \draw[thick, black!30, -stealth] (4.2, -1.5) -- node[right, text=black!30]{\circled{0}} ([xshift=-2mm]rps.south);
        \draw[thick, -stealth] ([xshift=-0.8cm]rps.south) -- node[left]{\circled{1}} (3.6, -1.5);
        \draw[thick, -stealth] (3.6, -2.5) -- node[left]{\circled{2}} ([xshift=-0.8cm]auth.north);
        \node[below of=auth, yshift=5mm] {\circled{3}};
        \draw[thick, -stealth] ([xshift=0.8cm]auth.north) -- node[right]{\circled{4}} (5.2, -2.5);
        \draw[thick, -stealth] (5.2, -1.5) -- node[right]{\circled{5}} ([xshift=0.8cm]rps.south);
        \node at ([xshift=1mm]rps.east) {\circled{6}};
    \end{tikzpicture}
    \caption[WebAuthn Authentication Workflow]{WebAuthn authentication workflow using a browser as client, cf.~\cite{webauthn}}
    \label{fig:webauthn_reg}
\end{figure}

 
Similarly, the login workflow starts by the user sending a passkey login request and the \ac{rp} generating a challenge. The \ac{rp} responds with the challenge and user information and the client requests a signature on the challenge from the authenticator through \ac{ctap}. The challenge is signed and is sent with user and device information to the \ac{rp}. The \ac{rp} locates the public key, through the sent information and verifies the signed challenge - the authentication is done. The authentication workflow is visualized in a more simplified way in Figure~\ref{fig:webauthn_reg}, using a browser as the client: {\footnotesize\circled{0}} the client starts passkey authentication, {\footnotesize\circled{1}} the \ac{rp} responds with a challenge, {\footnotesize\circled{2}} the challenge gets forwarded to the authenticator, {\footnotesize\circled{3}} the authenticator signs the challenge with the correct private key, {\footnotesize\circled{4}} then the authenticator sends the signed challenge to the client, {\footnotesize\circled{5}} \& {\footnotesize\circled{6}} the challenge gets forwarded to the \ac{rp}, gets verified and the authentication workflow is finished. 

\subsection{FIDO2}
\ac{fido} now incorporates both \ac{ctap} and WebAuthn.
\begin{quote}
    \textit{\ac{fido} standards use standard public key cryptography techniques to provide phishing-resistant authentication with cryptographic key pairs called passkeys. \ac{fido} is designed from the ground up to protect user privacy and prevent phishing. Every passkey is unique and bound to the online service domain. The protocols do not provide information that can be used by different online services to collaborate and track a user across the services. Biometric information, if used, never leaves the user’s device.}~\cite{fido2026}
\end{quote}
The entire workflow is already described in the sections above. In a more high level perspective:
\begin{enumerate}
    \item A user is visiting a website and registers.
    \item The device of the user is generating a passkey.
    \item The passkey is used to login into the webservice.
\end{enumerate}
This workflow is passwordless and also called \ac{uaf}. Another option is the \ac{u2f}, which incorporates a physical second factor, such as a (physical) security key.

\subsection{Threat Model}\label{subsec:thread_model}
The threat model assumes an attacker being able to compromise a victim's device with malware and perform \ac{aitm} attacks, such as a combination of \ac{aitm} and \ac{dns} Spoofing~\cite{spoofing}. Additionally, the victim may follow phishing links provided by the adversary. The user may use a synchronized authenticator, meaning that the passkeys are stored in the cloud, or a device-bound authenticator (local storage). The threat model focuses on attacks against end users rather than the \ac{fido} protocol itself, which is assumed to be cryptographically secure. The main goal of the adversary is to gain access to the user account or to pull out personal information.\par 
For the \ac{aitm} attack, a scenario would include the adversary to be in the same network as the victim or have malware installed which performs an \ac{aitm} against, e.g., an \ac{hsk}.\par 
The malware scenario includes the victim having malware installed on their device. This could be through a malicious download, a malware installation by the adversary locally (physical access) or a local malware installation by the victim via, e.g., a bad USB. In general, there are several possibilities.\par 
A phishing scenario, includes the attacker being able to trick the victim to click on a link and to trust the phishing website through visual deception. The attacker is also able to forward \ac{mfa} requests, like a text message verification on a login.

\section{Related Work}\label{sec:related_work}
This section summarizes attacks on \ac{fido} which have been shown in previous publications.\par

Li et al.~\cite{auth_rebind_attack} proposed an Authenticator Rebind Attack on the \ac{uaf} protocol, targeted at Android phones. The lack of effective entity authentication in the \ac{uaf} protocols allows an attacker to bind the victim's identity to the attacker's authenticator. This attack requires pre-installed malware on the victim's device, which redirects some of the traffic to the attacker or the attacker has to have some access prior to the attack. When the victim tries to enable fingerprint authentication, the request is forwarded to the attacker's authenticator. The attacker performs their own fingerprint verification initiated by the misused authenticator and the victim performs a fake fingerprint verification initiated by the malware. The authors demonstrated the attack's effectiveness against real-world applications, providing a real threat under the given circumstances.\par 
Barbosa et al.~\cite{barbosa} showed two possible attacks on \ac{fido}, one being an impersonation attack and the other one being a rogue key attack. On the impersonation attack, the adversaries capture the \textit{pinToken} by using an \ac{aitm} attack. The pinToken, a cryptographic key, is used by the client to send \ac{mac}-authenticated credential and assertion requests to the authenticator. The adversary intercepts active communication between the client and authenticator during the key agreement phase. During this phase, the adversary agrees with their own generated keys to the key agreement protocol. By doing that, the adversary, knowing the shared secret, creates a session with the client and authenticator. Now the adversary is able to create assertion requests to any relying party. Similar to Li et al.~\cite{auth_rebind_attack}, the second attack includes registering a rogue key on the \ac{rp}, but through the USB HID mechanism of the latest \ac{fido} version.\par 
Kumar Yadav et al.~\cite{local_attack}, present local attacks on the \ac{fido} protocol.
The attack methods are similar to the attacks presented previously~\cite{barbosa}, since one of the goals is to impersonate the victim. Here, browser extensions are used for performing an \ac{aitm} attack between the authenticator (a hardware security key) and the \ac{rp} through WebAuthn. The other goal is to bypass the cloning detection algorithm of an \ac{hsk} by \ac{fido}. The authors underline, that the \acp{rp} are lacking detailed error messages, to notify victims that they are being attacked.\par 
Mahdad et al.~\cite{deception_attack}, propose an attack which includes running malware in the user space and uses a browser extension. The malware detects keystrokes and mouse movements - the primary function is to capture username and password in an \ac{mfa} scenario. A hidden browser is launched in the background, navigating to the same service the user is trying to authenticate to. The browser extension, on the victim's browser, redirects the page to a similar-looking attacker-controlled page, prompting the victim to access their security key. Next, the hidden browser attempts to login, receives the challenge and forwards it to the attacker-controlled page. After the user authenticates with the key, the attacker is able to establish a hidden session.\par 
Donghyun et al.~\cite{hipass}, present a different kind of attack, on which the victim does not have to be preloaded with some kind of malware. This attack exploits the Bluetooth capabilities of \ac{ctap} and uses spoofing to reroute the victim to a phishing server. Upon visiting the phishing website and trying to log in, the adversary goes to the real \ac{rp} and attempts to login. The \ac{rp} responds with the relevant data needed to generate a QR-Code and the challenge. The adversary generates a legitimate QR-Code out of the received data and presents it to the victim. Next, the victim scans the QR-Code and tries to establish a Bluetooth connection to the device that generated the QR-Code. After the connection establishment, the challenge gets signed, forwarded to the adversary and thus logging the adversary in.

\section{Methodology}\label{sec:methodology}
This section enumerates attack vectors against \ac{fido} and outlines the requirements for successful exploitation. It is important to note that the following attacks do not target \ac{fido} directly, as it is assumed to be secure (\autoref{subsec:thread_model}); rather, they illustrate potential attack vectors, which further demonstrate the robustness of \ac{fido}. Each attack vector is classified according to its complexity level and setup effort of the attack; with easy, medium or hard/impossible. A key observation is that all attack scenarios require substantial effort. In all cases, an attacker must either compromise the victim's system or be within close physical range. Notably, none of the described attacks offer a method for remotely (high range) exploiting \ac{fido} without first compromising the victim's system.\par 
As already mentioned in the threat model (\autoref{subsec:thread_model}), the main goal of an attacker is to gain access to the user account or to gather personal information. Personal user information could then be used to open new attack vectors on the victim; this won't be further discussed. For accessing the account, several paths exist like the ones mentioned in \autoref{sec:related_work} - those will be included below.

\subsection{Getting Victim's Key}
One path, which will be practically impossible, especially on synchronized authenticators (due to cloud storage), is to get the private key for an existing account. With malware, it is possible to extract private keys, depending on where they are stored. 
Password managers like KeePassXC~\cite{keepassxc2026} store the passkeys in an encrypted database on the file system.
To pull out the private keys, the encryption of the database has to be cracked; this requires malware to be installed or local access to the target device, which can be hard. Extraction methods vary from different authenticators, of course; e.g., \ac{hsk}s have to be locally accessed since the keys are generated and stored on the device. The private key does not leave the \ac{hsk} but is only used for signing, making it significantly harder to extract.

\subsection{Victim Impersonation}\label{subsec:victim_impersonation}
Other paths are more feasible, like impersonating the victim. For this to work, the authentication process of \ac{fido} has to be hijacked.
One option would be to create a session with the \ac{rp} and the client in the key agreement phase, which requires an \ac{aitm} between the authenticator and the client. Barbosa et al.~\cite{barbosa} showed this practically with the USB HID transport mechanism; due to lack of authentication on \ac{ecdh} in \ac{ctap}, an attacker is able to create a session to the \ac{rp}. The \ac{aitm} attack requires malware to be installed on the victim's device. If the attacker is positioned between the authenticator and the client, a session can be established or the attacker's own key can be registered on the \ac{rp}. It is important to mention that the success of this attack depends on the user being present and on malware being installed. Due to this, this path of the impersonation attack would be hard.

\subsection{Bluetooth \ac{aitm}}
The aforementioned Bluetooth attack~\cite{hipass}, where the victim does not have to be preloaded with malware, operates in a similar way. This attack adds the use of a phishing page, which is only there for visual deception. The authentication is performed on an attacker's remote device. Further, the attack requires the attacker to be in range of a user, because the Bluetooth authentication of \ac{ctap} has to be hijacked. This path depends on a user wanting to authenticate with QR-Code scanning, the attacker being in range and a successful phishing attack. This attack will be classified as medium, because the user does not have to be preloaded with malware but still has to interact with a phishing page. 

\subsection{Authenticator Deception}\label{subsec:authenticator_deception}
Passkeys are phishing resistant~\cite{fido2026}, but still phishing websites can be utilized to gain access to the victim's account. To login as the victim, the attacker has to own a website, with the same domain as the \ac{rp}. Since this is not possible, the attacker performs a DNS spoofing attack~\cite{spoofing}, forwarding requests to the phishing website. A problem which arises is the SSL certificate, which the attacker does not own. Two sub paths arise, mainly using an insecure connection or forcing the victim to trust a self signed certificate. An HTTP connection might technically work, most browsers don't explicitly warn you as they would with self-signed certificates - for inattentive users this would go unnoticed. However, the in-browser WebAuthn API only allows HTTP access when running on localhost, making it impossible to use WebAuthn over an HTTP connection in a real attack scenario.

When using HTTPS the victim has to trust the self signed certificates, because it is not possible to gain the real certificate of the \ac{rp}. This requires malware or physical access to the victim's device to install the certificate, either into the browser or into the OS certificate authorities. The latter option would be even harder, because this path usually needs administrative rights. First, a DNS spoof is performed and the victim is lured to the phishing website of the \ac{rp}. Since the interface and the domain is the same as in \ac{rp}, the authenticator will recognize it and start the process. In the background, the attacker starts a session with the real \ac{rp} and forwards the requests, including the challenge, to the phishing website. After the victim's authenticator has signed the challenge, it is forwarded back to the attacker, allowing login access. To remove any suspicions, the phishing server shows an error and recommends reloading the page, so the victim will be forwarded to the real \ac{rp}. This requires the attacker to turn off DNS spoofing, after the successful login. The setup of this attack is easy to realize but access to the target's device is required, making the attack medium to hard.

\subsection{Registering Attacker's Passkey}\label{subsec:reg_attack_passkey}
Similar to the victim impersonation attacks, an attacker can try registering their own passkey, on behalf of the victim.
One way would be to hijack the process of adding passkeys on the \ac{rp}. To motivate the victim, a fake email can be sent, which recommends that the user add passkeys. When the victim now tries to add passkeys, the attacker has to sit in between the authenticator and the client, thus needing an \ac{aitm}. Since the goal is to register the keys on the real \ac{rp}, a simple \ac{aitm} over the network and DNS spoofing is not sufficient. The \ac{aitm} has to be run locally, on the victim's device, which again requires malware.
When a request to add passkeys is observed, the challenge is signed with the attacker's own private key, and the victim's public key is replaced. This attack is not too stealthy, since the whole \ac{fido} operation fails on the victim's side. To optimize this, the attacker can make it look as if the operation was successful by forwarding and modifying packets, marking the authentication as successful from the \ac{rp}'s side. Obviously, the victim won't be able to login with their newly created passkey, due to the public key not matching. The described method will be ranked as hard, because malware has to be installed. The second option for registering own passkeys is to login as the user via traditional methods like email and password. For this, a phishing website is used which captures the login data.
After the victim enters their credentials, the attacker logs in and registers attacker-controlled keys. This attack is easy to realize and will also be ranked in this category. This attack succeeds only if the victim uses email and password authentication. If a passkey is used as a second factor, this attack fails, because the phishing domain is not matching. Here, a phishing website as proposed in \autoref{subsec:authenticator_deception}, can be used to hijack the \ac{ctap} communication. This attack focuses on the attacker registering their own keys~-- the phishing website is used as a tool to gain initial access. In case of password change, the key remains valid. Unlike conventional phishing attacks, which lose effectiveness once the victim changes their password, this attack establishes persistent access through a registered passkey that remains valid independently of any subsequent credential changes.

\subsection{Passkey Reduction}\label{subsec:passkey_reduction}
Many websites~\cite{passkeys2026} provide the option to register a passkey instead of typical password authentication. But password authentication is not gone and is still used. This enables a user to authenticate via multiple methods. Since one of the main goals was to gain access to the account, an attacker can utilize a passkey reduction attack. First, the victim is lured to a phishing page and goes to the login page. The passkey option will either not be shown or upon choosing \textit{passkey authentication}, the fake \ac{rp} signals an error and prompts the victim for password authentication. This attack depends on the user believing that passkey authentication is currently unavailable. The setup, on the other hand, is easy to realize and thus the attack is ranked in the easy category. When passkeys are configured as a second factor in \ac{mfa}, the attacker has to spoof the real \ac{rp}'s domain to sign the challenge as described in \autoref{subsec:authenticator_deception}.

\subsection{Infected Authenticator}\label{subsec:infected_authenticator}
The final attack on passkeys, where the goal is to gain access to the victim's account, uses an \textit{infected authenticator}. The attacker modifies the underlying source code of an authenticator, making it generate known key pairs. When a user creates accounts on the \ac{rp}, the attacker will be able to login, since the private key is known. Either the application's source code or the used cryptographic libraries, which are used to generate the keys, can be modified. Manipulating system libraries can be more challenging because it typically requires superuser permissions. When a library is modified, all applications using the library will generate known keys. The complexity in this attack lies in the installation of malware (infected authenticator) or malware which modifies the target library or application. This attack is limited to local authenticators, meaning the key pairs must be generated locally on the victim's device. This attack path is ranked medium to hard, because of malware installation but the easy realisation. The complexity increases even more when system libraries get modified. However, implementing malicious code into applications is straightforward. Alternative approaches include exfiltrating generated keys over the network. A downside is that such behaviour can be detected by anti-malware systems~\cite{antivirus}; e.g., an anti-malware system can detect and flag the use of sockets, because it can associate it to network activity. Other malware-reliant attacks described in this section, such as local AITM or certificate store manipulation, do not require outbound network connections to attacker-controlled infrastructure and are therefore less likely to trigger such detection mechanisms.

\subsection{RP Impersonation}
In the final attack scenario, an attempt is made to extract personal information from the victim by causing the victim to believe that authentication is being performed with a legitimate \ac{rp}.
To achieve this, an attacker must spoof both the \ac{rp}'s website and the \ac{ctap} interaction. As a result, the victim is led to believe that a successful login has occurred, even though the entire process is a visual deception. This attack requires the attacker to replicate the genuine user interface of a logged in RP session. Further, the \ac{fido}-related library on the victim's device must be modified so that the authentication process appears successful regardless of the actual input. Once the victim is “logged in”, the attacker can display prompts requesting personal information under the guise of security checks. Of course, there are numerous strategies of how information can be extracted. A simpler variant is to force the victim to bypass passkey authentication entirely and directly request personal information, similar to \autoref{subsec:passkey_reduction}. This variant only requires a successful phishing attack, making it relatively easy to execute. Note that this simpler variant does not rely on FIDO2 at all and would be equally effective against any authentication method; it is included here because it represents a degraded form of the full \ac{rp} impersonation attack, which requires the installation of malware and the manipulation of \ac{rp} libraries, making it significantly more complex and thus ranked hard.

Table~\ref{tab:attacks_overview} provides an overview of the attacks characteristics, which will be analysed in the next section. In the first row of the table, IE is used as an abbreviation for \textit{Infected Authenticator} and AD for \textit{Authenticator Deception}.
\begin{table}[h!]
    \centering
    \rowcolors{2}{black!10}{white}
    \begin{tabular}{lcc}
        \toprule
        \textbf{Criterion} & \textbf{IE} & \textbf{AD} \\
        \midrule
        Stealthiness & High & Medium \\
        Feasibility & Medium & High \\
        Victim Interaction & Low & High \\
        Time Consumption & Low & Low-High \\
        Privileges & Low-High & Low-High \\
        \bottomrule
    \end{tabular}
    \caption{Overview of the attacks}
    \label{tab:attacks_overview}
\end{table}

\section{Experiments}\label{sec:experiments}
Two experiments were conducted to demonstrate: first, that executing an attack on \ac{fido} is highly resource-intensive, and second, that the attacks closely resemble phishing methods. Since passkeys cannot be phished directly, \autoref{subsec:exp_infected_authenticator} describes an attacker being able to possess the same passkey. Similarly, \autoref{subsec:exp_authenticator_deception} mirrors a typical phishing scenario, where a legitimate website is displayed, but the attacker's goal is to steal the proof of authentication.

\subsection{Infected Authenticator}\label{subsec:exp_infected_authenticator}
The first experiment follows the method described in \autoref{subsec:infected_authenticator}. In this attack, KeePassXC, a local authenticator, is used, and only the source code of the binary is modified; no changes are made to the system libraries responsible for \ac{fido}. As previously mentioned, the source code of the authenticator is altered so that an attacker can obtain the same key that the victim generates during the registration process. KeePassXC relies on the Botan cryptographic library, which is used to generate asymmetric keys. Specifically, KeePassXC requires the generation of \ac{ecdsa}, \ac{rsa}, and \ac{ed25519} keys.\par 
The attack proceeds as follows: the attacker first generates their own public-private key pairs using Botan, which are saved in PEM format for later import into the KeePassXC database. This key is then embedded directly into KeePassXC’s source code as a string variable. In the modified code, the private key in PEM format is loaded into the appropriate object type, while the rest of the code remains unchanged. This ensures that every time a new passkey is registered, the same private key is used, leading to the same public key being generated by Botan.\par 
Depending on the \ac{rp}, the public key (used to verify the challenge) is associated with a unique credential-ID. KeePassXC generates these credential IDs, which are then stored on the \ac{rp}. Therefore, the attacker must also know the correct credential-ID, which is also easily embedded into the source code. The binary is then placed on the victim's computer, essentially replacing the old KeePassXC binary. After the victim registers their passkey on the \ac{rp}, the attacker can authenticate and log in using the previously generated key pair. This experiment was conducted on a Firefox browser, specifically for logging into a Google account.

\subsection{Authenticator Deception}\label{subsec:exp_authenticator_deception}
The second experiment realizes the Authenticator Deception attack, which was described in \autoref{subsec:authenticator_deception}. The attack setup was as follows. The victim was using a Windows 10 machine, while the attacker operated from a Kali Linux system. The victim had a registered test account on the target domain: \textit{linear.app} -- in general, any domain can be targeted, provided it implements \ac{fido}. The attacker also possessed a valid account on linear.app, which was required in order to modify authentication requests dynamically during the attack.\par 
First, the attacker generated a valid TLS certificate for the fraudulent frontend server that would later be presented to the victim as a result of DNS spoofing~\cite{spoofing}. Since modern browsers do not trust self-signed certificates by default, the attacker requires prior access to the victim’s machine. Specifically, the attacker created a custom \ac{ca} and signed the frontend certificate with this \ac{ca}, allowing the certificate to be valid. The custom \ac{ca} certificate was then imported into the victim’s browser (Firefox) trust store, effectively forcing the browser to trust the attacker’s frontend certificate. 
Additionally, the frontend certificate contained a \ac{san} extension as well as a Common Name (CN) matching the target domain, in this case linear.app. Only after these preparatory steps was the victim’s machine considered compromised and the actual attack could begin.\par 
For the attacker-controlled infrastructure, an NGINX web server was used as the frontend, combined with an Express-based Node.js application acting as the backend. This backend handled incoming requests from the victim’s browser and proxied or modified them as necessary before forwarding them to the legitimate \ac{rp}.\par 
The attacker initiated an ARP spoofing attack~\cite{spoofing}, thereby poisoning the victim’s ARP table and replacing the legitimate router’s MAC address with the attacker’s own. In parallel, a DNS spoofing attack was launched that mapped all DNS requests for linear.app and *.linear.app to the attacker’s local IP address. As a result, all traffic intended for linear.app was redirected to the attacker-controlled frontend server.\par 
When the victim navigated to linear.app using the compromised browser, the request was routed to the attacker’s NGINX frontend {\footnotesize\circled{1}}. The cloned website provided the option to authenticate using passkeys, which the victim selected. Upon initiating passkey authentication, a challenge was required. To obtain a valid challenge, the attacker simultaneously initiated an authentication attempt with the legitimate linear.app \ac{rp} {\footnotesize\circled{2}}. Using Burp Suite as an \ac{aitm} proxy, the attacker intercepted the authentication request and extracted both the challenge and the associated authenticationId {\footnotesize\circled{3}}. This identifier is a temporary value that uniquely binds a specific authentication session and is required to complete the passkey authentication flow.\par 
The attacker’s backend then forwarded the captured challenge to the cloned frontend, prompting the victim to sign it using their passkey {\footnotesize\circled{4}} . In this setup, the authenticator software used by the victim was Bitwarden~\cite{bitwarden2026}. Once the victim approved the request {\footnotesize\circled{5}}, the signed challenge was produced and returned to the attacker-controlled backend {\footnotesize\circled{6}}.\par 
At this stage, the attacker authenticated to the legitimate relying party using their own passkey {\footnotesize\circled{7}}. Before completing the authentication, the attacker manually modified the final request by replacing their own signed challenge and user identifier with those obtained from the victim. By submitting the victim’s signed challenge together with the victim’s user ID, the attacker was able to successfully authenticate as the victim on the real linear.app service.\par 
All steps described above were performed manually. The attacker manually transferred the challenge, initiated authentication flows, and edited the final authentication request.
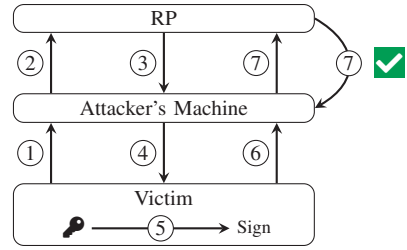
\begin{figure}[htb!]
    \centering
    \begin{tikzpicture}
        \footnotesize
        \node[anchor=south, draw, minimum width=4cm, rounded corners] (rp) at (4.4, -1) {RP};
        \node[draw, minimum width=4cm, rounded corners] (attacker) at (4.4, -2) {Attacker's Machine};
        \node[anchor=north, draw, minimum width=4cm, rounded corners, text width=3cm, text centered] (victim) at (4.4, -3) {Victim\\\Large\vphantom{j}};
        \node (key) at (3.2, -3.6) {\faKey\vphantom{j}};
        \node (sgn) at (5.6, -3.6) {\scriptsize Sign};
        \draw[thick, -stealth] ([xshift=-1.5cm]victim.north) -- node[left]{\circled{1}} ([xshift=-1.5cm]attacker.south);
        \draw[thick, -stealth] ([xshift=-1.5cm]attacker.north) -- node[left]{\circled{2}} ([xshift=-1.5cm]rp.south);
        \draw[thick, -stealth] (rp.south) -- node[left]{\circled{3}} (attacker.north);
        \draw[thick, -stealth] (attacker.south) -- node[left]{\circled{4}} (victim.north);
        \draw[thick, -stealth] (key.east) --node{\circled{5}} (sgn.west);
        \draw[thick, -stealth] ([xshift=1.5cm]victim.north) -- node[left]{\circled{6}} ([xshift=1.5cm]attacker.south);
        \draw[thick, -stealth] ([xshift=1.5cm]attacker.north) -- node[left]{\circled{7}} ([xshift=1.5cm]rp.south);
        \draw[thick, -stealth] (rp.east) to[out=-30, in=30, looseness=1.5] node{\circled{7}} (attacker.east);
        \fill[green!90!black] (7.2, -1.2) rectangle (7.6, -1.6);
        \node[text=white] at (7.4, -1.4) {\normalsize\faCheck};
    \end{tikzpicture}
    \caption[AD Flow]{Authenticator Deception Flow}
    \label{fig:ad_flow}
\end{figure}

\section{Results and Discussion}\label{sec:results_and_discussion}
Both experiments succeeded as described; however, several important nuances were observed.\par 
For the authenticator deception attack, it was important that the DNS cache on the victim’s machine was flushed. If stale DNS entries were present, the victim’s browser would continue to resolve the domain to the legitimate \ac{rp}, bypassing the attacker-controlled server. Since compromising or gaining access to the victim’s machine was already a prerequisite for the attack, the attacker was able to flush the DNS cache manually. In a realistic malware scenario, this step is straightforward to automate, as flushing the DNS cache requires only a simple system command that typically does not require elevated privileges. This ensured that subsequent DNS queries were forwarded to the DNS server where the attacker was positioned as an \ac{aitm} and could respond with their own IP address, pointing to the malicious frontend server. Clearly, in a real-world attack scenario, such an attack would not be executed in such a manual fashion. Instead, automation tools would be employed, such as Selenium for browser automation and a programmable AITM proxy capable of modifying packets dynamically. For demonstration purposes, we did not create a visual clone of the legitimate website; rather, the malicious frontend only provided an option to authenticate via passkeys. In a realistic attack scenario, however, the website would need to be visually indistinguishable from the legitimate \ac{rp}'s website to avoid raising suspicion on the victim’s side. Finally, it is important to emphasize that this attack is not trivial to carry out. A critical requirement is the compromise of the victim’s machine, which either involves tricking the user into installing malware that modifies the browser’s trust store, flushes the DNS cache, and performs related actions, or requires physical access to the system in order to apply these modifications manually. Additionally, this attack is highly targeted at a specific individual, as the attacker must know in advance which website to clone, be located within the same network as the victim and successfully perform an ARP spoofing attack~\cite{spoofing} against the target. The attack succeeds only if all these conditions are met. Moreover, the attack will fail if either the router or the victim’s machine implements anti-ARP-spoofing mechanisms, which would effectively prevent DNS spoofing.\par 
The infected authenticator attack targeted a specific local authenticator, namely KeePassXC. This attack does not apply to cloud-based authenticators such as Bitwarden. Consequently, the victim was required to use an authenticator that generates cryptographic keys locally. Under these conditions, malware can hijack the key-generation process and deliberately produce attacker-controlled or predictable keys. The described method demonstrated that previously generated keys were hardcoded as strings directly into the source code of the authenticator. This approach causes the authenticator to always generate the same keys and the same credential-ID during passkey registration. As a result, a victim using an infected authenticator would only be able to generate a single passkey per \ac{rp}. A more effective approach from an attacker’s perspective would be to use a predictable algorithm that generates keys and credential-IDs following a known pattern, or to rely on a predictable seed value from which keys are derived. Alternatively, multiple keys and credential-IDs could be embedded into the program, although this approach remains limited. Another option discussed would be to leave the authenticator’s behavior unchanged while sharing the generated keys and credential-IDs over the network to the attacker. This would allow the attacker to obtain valid credentials without immediately breaking the expected functionality of the authenticator.\par 
The demonstrated attacks showed that executing an attack on \ac{fido} is time‑consuming and therefore resource‑intensive.

\section{Conclusion and Future Work}\label{sec:Conclusion_and_Future_Work}
While passwords remain a significant security weakness in modern systems, this work demonstrates that attacking \ac{fido}-based authentication is far from trivial. Our experiments and analysis show that successfully compromising such systems is complex and time-consuming. In addition, multiple strict prerequisites must be fulfilled, such as physical proximity to the victim or prior compromise of the victim’s device. These constraints significantly raise the bar for attackers and result in a threat model that is considerably more robust than traditional password-based authentication.\par 
This work analysed possible attack vectors against \ac{fido} authentication and demonstrated that \ac{fido} largely fulfills its security promises. In particular, passkeys cannot be phished using conventional techniques since the private key never leaves the authenticator. As a result, many well-known attacks that are effective against password-based systems are rendered ineffective.\par 
Future work includes the automation of the authenticator deception attack, as well as research into attack techniques that could target \ac{fido} authentication without relying on device compromise or requiring an attacker to be within a specific physical range of the victim.

\section*{Acknowledgements}
This research was conducted by Alexander Berladskyy as part of a Computer Science Research Project at Kiel University of Applied Sciences.

\balance
\printbibliography

@article{rsa,
author = {Rivest, R. L. and Shamir, A. and Adleman, L.},
title = {A method for obtaining digital signatures and public-key cryptosystems},
year = {1978},
issue_date = {Feb. 1978},
publisher = {Association for Computing Machinery},
address = {New York, NY, USA},
volume = {21},
number = {2},
issn = {0001-0782},
url = {https://doi.org/10.1145/359340.359342},
doi = {10.1145/359340.359342},
journal = {Commun. ACM},
month = 2,
pages = {120–126},
numpages = {7}
}

@article{ctap,
  title="Client to Authenticator Protocol {(CTAP)}",
   author = "Brand, Christiaan and Czeskis, Alexei and Ehrensvard, Jakob and Jones, Michael B. and Kumar, Akshay and Lindemann, Rolf and Powers, Adam and Verrept, Johan",

  year={2019},
  month=1,
 url={https://fidoalliance.org/specs/fido-v2.0-ps-20190130/fido-client-to-authenticator-protocol-v2.0-ps-20190130.html},
}

@misc{fido_security,
      author = {Nina Bindel and Cas Cremers and Mang Zhao},
      title = {{FIDO2}, {CTAP} 2.1, and {WebAuthn} 2: Provable Security and Post-Quantum Instantiation},
      howpublished = {Cryptology {ePrint} Archive, Paper 2022/1029},
      year = {2022},
      url = {https://eprint.iacr.org/2022/1029}
}

@article{webauthn,
  title = "Web Authentication: An {API} for accessing Public Key Credentials Level 2",
  author = "Jeff Hodges and J.C. Jones and Michael B. Jones and Akshay Kumar and Emil Lundberg and Dirk Balfanz and Vijay Bharadwaj and Arnar Birgisson and Alexei Czeskis and Hubert Le Van Gong and Angelo Liao and Rolf Lindemann",
  year = {2021},
  month = 4,
  url = {https://www.w3.org/TR/webauthn-2/},
}

@ARTICLE{auth_rebind_attack,
  title     = "Authenticator Rebinding Attack of the {UAF} protocol on mobile
               devices",
  author    = "Li, Hui and Pan, Xuesong and Wang, Xinluo and Feng, Haonan and
               Shi, Chengjie",
  journal   = "Wirel. Commun. Mob. Comput.",
  publisher = "Hindawi Limited",
  volume    =  2020,
  pages     = "1--14",
  month     =  9,
  year      =  2020,
  language  = "en"
}

@inproceedings{barbosa,
author = {Barbosa, Manuel and Cirne, Andr\'{e} and Esqu\'{\i}vel, Lu\'{\i}s},
title = {Rogue key and impersonation attacks on FIDO2: From theory to practice},
year = {2023},
isbn = {9798400707728},
publisher = {Association for Computing Machinery},
address = {New York, NY, USA},
url = {https://doi.org/10.1145/3600160.3600174},
doi = {10.1145/3600160.3600174},
booktitle = {Proceedings of the 18th International Conference on Availability, Reliability and Security},
articleno = {14},
numpages = {11},
location = {Benevento, Italy},
series = {ARES '23}
}

@ARTICLE{hipass,
  author={Kim, Donghyun and Shin, Junseok and Ryu, Gwonsang and Choi, Daeseon},
  journal={IEEE Access}, 
  title={HiPass: Hijacking CTAP in Passkey Authentication}, 
  year={2025},
  volume={13},
  number={},
  pages={92086-92101},
  doi={10.1109/ACCESS.2025.3570377}}

@misc{local_attack,
      title={A Security and Usability Analysis of Local Attacks Against FIDO2}, 
      author={Tarun Kumar Yadav and Kent Seamons},
      year={2023},
      eprint={2308.02973},
      archivePrefix={arXiv},
      primaryClass={cs.CR},
      url={https://arxiv.org/abs/2308.02973}, 
}

@inproceedings{deception_attack,
author = {Mahdad, Ahmed Tanvir and Jubur, Mohammed and Saxena, Nitesh},
title = {Breaching Security Keys without Root: FIDO2 Deception Attacks via Overlays exploiting Limited Display Authenticators},
year = {2024},
isbn = {9798400706363},
publisher = {Association for Computing Machinery},
address = {New York, NY, USA},
url = {https://doi.org/10.1145/3658644.3690286},
doi = {10.1145/3658644.3690286},
booktitle = {Proceedings of the 2024 on ACM SIGSAC Conference on Computer and Communications Security},
pages = {1686–1700},
numpages = {15},
location = {Salt Lake City, UT, USA},
series = {CCS '24}
}

@book{antivirus,
author = {Szor, Peter},
title = {The Art of  Computer Virus Research and Defense},
year = {2005},
isbn = {0321304543},
publisher = {Addison-Wesley Professional},
}

@article{spoofing,
  title={A comprehensive analysis of spoofing},
  author={Babu, P Ramesh and Bhaskari, D Lalitha and Satyanarayana, CH},
  journal={International Journal of Advanced Computer Science and Applications},
  volume={1},
  number={6},
  year={2010},
  publisher={Science and Information (SAI) Organization Limited}
}

@misc{phishing_report,
  author       = {{APWG}},
  title        = {Phishing Activity Trends Report},
  year         = {2025},
  month        = {12},
  note         = {Activity: July-September 2025, Published: 9 December 2025},
  howpublished = {APWG Report Q3 2025},
}

@misc{yubico,
  author       = {{Yubico}},
  title        = {Yubico: Security Keys and Authentication Solutions},
  howpublished = {\url{https://www.yubico.com/}},
  note         = {Accessed: 2026-03-14},
  year         = {2026},
}

@misc{fido2026,
  author       = {{FIDO Alliance}},
  title        = {FIDO Specifications},
  howpublished = {\url{https://fidoalliance.org/specifications/}},
  note         = {Accessed: 2026-03-14},
  year         = {2026}
}

@misc{mozillaWebAuthn2026,
  author       = {{Mozilla Developer Network}},
  title        = {Web Authentication API (WebAuthn)},
  howpublished = {\url{https://developer.mozilla.org/en-US/docs/Web/API/Web_Authentication_API}},
  note         = {Accessed: 2026-03-14},
  year         = {2026}
}

@misc{keepassxc2026,
  author       = {{KeepassXC Team}},
  title        = {KeePassXC: Cross-Platform Password Manager},
  howpublished = {\url{https://keepassxc.org/}},
  note         = {Accessed: 2026-03-14},
  year         = {2026}
}

@misc{passkeys2026,
  author       = {{Passkeys.io}},
  title        = {Who Supports Passkeys},
  howpublished = {\url{https://www.passkeys.io/who-supports-passkeys}},
  note         = {Accessed: 2026-03-14},
  year         = {2026}
}

@misc{bitwarden2026,
  author       = {{Bitwarden}},
  title        = {Bitwarden Password Manager},
  howpublished = {\url{https://bitwarden.com}},
  note         = {Accessed: 2026-03-14},
  year         = {2026}
}

\end{document}